# Photoferroelectric oxides

I. Fina[a,*], C. Paillard[b], B. Dkhil[c]

[a]Institut de Ciència de Materials de Barcelona (ICMAB-CSIC), Campus UAB, Bellaterra 08193, Barcelona, Spain
[b]Physics Department, University of Arkansas, Fayetteville AR 72701, USA.
[c]Laboratoire Structures, Propriétés et Modélisation des Solides, CentraleSupélec, CNRS-UMR 8580, Université Paris-Saclay, 91190 Gif-sur-Yvette, France

**Abstract**

Giant photovoltaic effect due to bulk photovoltaic effect observed in multiferroic BiFeO$_3$ thin films has triggered a renewed interest on photoferroelectric materials for photovoltaic applications. Tremendous advance has been done to improve power conversion efficiency (up to up to 8.1%) in photoferroelectrics via absorption increase using narrow bandgap ferroelectrics. Other strategies, as it is the more efficient use of ferroelectric internal electric field, are ongoing. Moreover, as a by-product, several progress have been also achieved on photostriction that is the photo-induced deformation phenomenon. Here, we review ongoing and promising routes to improve ferroelectrics photoresponse.

## 1. Introduction

Ferroelectric materials show spontaneous electric polarization switchable by the application of an external electric field. In photoferroelectric materials, most of them oxide materials, light affect the electric or polar properties of the ferroelectric material. Photoferroelectric materials in thin film form can show giant open circuit voltage under illumination [1], or switchable photoelectric response [2-4], which are exclusive characteristics that make them interesting for applications. Recently, in bulk photoferroelectrics, another interesting property, photostriction, has been shown to be high [5]. Photovoltaic (among others) applications of photoferroelectrics materials [6-10] is a research area receiving an in crescendo interest due to the promise of high photovoltaic efficiencies [11]. Here, we will focus on the last advances on the study of photoferroelectric oxide thin films focusing on these results that try to help on the improvement of photovoltaic as well as photostriction efficiency.

## 2. State of the art

The main factors that determine the efficiency in photoferroelectrics are those that also apply in any photovoltaic device: **absorption**, **charge collection**, and **exciton dissociation** (Fig.1). The main efforts done by the scientific community to increase the **absorption** mainly lie on the narrowing of the investigation of narrow bandgap ferroelectrics: solid state solutions ($[KNbO_3]_{1-x}[BaNi_{1/2}Nb_{1/2}O_{3-\delta}]_x$ [12]), novel ferroelectric materials ($Bi_2FeCrO_6$ with record efficiency of 8.1% [13]) or other [14]. Hexagonal manganites has been also proposed as a potentially interesting material [15], due to its small band gap (near 1.5eV); experimental results are still scarce but showing improved efficiency compared with the archetypical ferroelectrics [16].

**Charge collection efficiency** is related to the **internal electric fields (IEF)** presence, although it also depends on carriers mobility and lifetime. In photoferroelectrics, the charge collection can also be mediated by the **bulk photovoltaic effect (BPV)**. In the BPV the photocurrent steams from the asymmetric distribution of the momentum of the photogenerated carriers due to the noncentrosymmetric crystal environment [17]; although it can be also explained in terms of a virtual shift in the real space [18, 19]. Seminal examples of BPV are doped $LiNbO_3$ [20], and $BaTiO_3$ [21]. More recently BPV has been reported in thin film form in $BiFeO_3$ [22-24], and $BaTiO_3$ [25-27]. Although BPV is responsible of giant open circuit voltages under illumination, the generated photocurrent is low, which is ascribed to the short distance ($l_0$=10-100nm) [17, 28] that photocarriers can travel. In BPV efficient collection has been achieved reducing the device size: by growing it in thin film form [25], or by the use of micro- or nano- meters size tips directly on the ferroelectric surface (tip-enhanced technique) [22], reaching high local efficiencies [26] although the device efficiency remains low [29]. Ferroelectric domain wall engineering has been also proposed to improve efficiency [1]; with contradictory results regarding its relevant role[23, 24]. The IEF in a ferroelectric material are the imprint electric field ($E_{imp}$) and the depolarization electric field ($E_{dep}$), and therefore $E_{int} = E_{dep}+E_{imp}$. $E_{imp}$ can be generated by strain gradients among other effects, which is particularly interesting because they can generate very strong electric fields (up to 2 MV/cm [30]). $E_{dep}$ results from the unscreened ferroelectric surface charge [31], and its upper limit is equal to the electric field generated by an absolutely unscreened polarization, (in $BaTiO_3$ this is $E_{dep} <$ 5MV/cm). However, $E_{dep}$ also induces ferroelectricity suppression (with the consequently $E_{dep}$ suppression) and one must found a tradeoff between ferroelectricity and $E_{dep}$. IEF in oxide ferroelectric thin films has been widely studied and the different contributions of $E_{dep}$ and $E_{imp}$ to the photocurrent have been analyzed finding conflicting results on their relative relevance [3, 32-37]. Also present in ferroelectric devices but not exclusive of ferroelectric materials is the electric field associated to the presence of Schottky barrier at the metal/ferroelectric interface, usually called built-in electric field ($E_{bi}$). The $E_{bi}$ presence and its dependence on polarization results in interesting features such as switchable rectifying mechanism [2, 38, 39], but in this case the efficiency is limited by the ferroelectric bandgap [10] and larger efficiencies than that found for semiconductor based devices cannot be expected. Note that artificial design of IEF, via for instance nanolayering (combination of pure ferroelectric material and with high absorption oxygen deficiency one) can promote the delocalization of electronic states and thereby significantly enhance the BPV output, as was recently suggested from first-principle calculations [40].

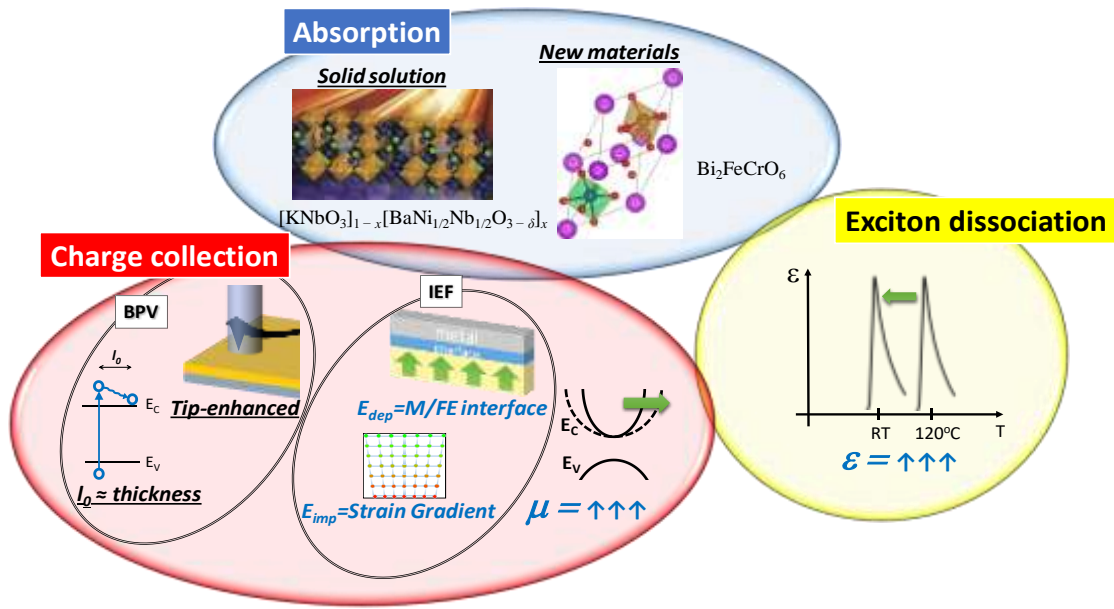

**Fig. 1**. Summary figure of the factors that determine photovoltaic efficiency and possible strategies to improve them using oxide ferroelectric materials. In black bold are the strategies that have found success, in blue are potentially interesting new strategies.

The ability of a material to change shape under illumination, called photostriction, represents an additional functional properties of ferroelectric materials that adds up to the already discussed photovoltaic effect. Photostriction can in principle appear in all materials, through light-induced thermal expansion, potential deformation (modification of the electron density distribution and thus the internal electronic pressure), or electrostriction [41]. Ferroelectric materials provide an additional handle with the natural occurrence of piezoelectricity, which coupled with a photo-induced electric field, creates a strain field under illumination. So far, photostriction in ferroelectrics has mainly been reported in classical ferroelectric materials such as $LiNbO_3$ [42], $(Pb_{1-x}La_x)(Zr_{1-y}Ti_y)O_3$ [42-45], and most recently research has focused on multiferroic $BiFeO_3$ [5, 46] for it has a relatively low bandgap (2.2-2.7 eV) compared with other archetypical ferroelectric material. Actual use of photostrictive materials as optically-driven mechanical actuators, such as micro-walking robots and photophones, were proposed long ago based on a PLZT bimorph structure [44]. More recent applications were also investigated by depositing a magnetic nickel film on top of a $BiFeO_3$ single crystal. It was demonstrated that the coercive field of the nickel magnetic hysteresis loop could be increased by 50% when the bismuth ferrite crystal was illuminated [47], thereby opening the way towards optically-controlled magnetic memories.

Recent studies have also focused on the ultra-short dynamics (below the picosecond or femtosecond) of the photostriction effect in thin films. In particular, below 100 picosecond strains were photo-induced in bismuth ferrite thin films, as monitored from Bragg peak shifts in time-resolved X-ray diffraction pump-probe experiments [48-50]. Similar results were observed in lead titanate thin films [51]. Also, transient reflectivity pump-probe experiments have shown the possibility to excite GHz acoustic waves, and in particular shear acoustic waves with a large intensity compared to longitudinal ones could be excited in bismuth ferrite ceramics and single crystals [52, 53].

## 2. Challenges and Future Prospects

The discovery of new materials with a bandgap near the ideal ≈1.5eV (lower limit to avoid thermalization [54]) is a very important accomplished milestone in photoferroelectrics. However, it is also important to look for new oxide materials focusing on the optimization of carrier mobility. In this direction cationic substitution is a potentially interesting strategy. For instance, if one takes the archetypical example of BaTiO$_3$, where the conduction band is mainly dominated by carriers mobbing through d-orbitals, appropriate doping can help on manipulating the conduction band to allow mobility through s-orbitals, thus enhancing it, resulting in a larger charge collection, and enhanced efficiency.

Other alternatives to improve the efficiency are those that include take more advantage of the intrinsic differential properties that ferroelectric materials show. These may be summarized in: IEF and high dielectric permittivity. Although there are several works exploring the ferroelectric IEF role on the phothovoltaic efficiency, there is lack of work focused on their manipulation to improve performance [55]. As mentioned, large internal electric fields can be achieved by the engineering of the electrode/ferroelectric interface to enlarge the $E_{dep}$ or by the generation of large strain gradients to improve the $E_{imp}$. Otherwise, the large dielectric permittivity near the Curie point in ferroelectrics, which can be easily tailored by epitaxial strain [56] or composition [57], can help on the **exciton dissociation**, which binding energy is inversely proportional to dielectric permittivity. Also selection of appropriate electrodes might help not only to improve the IEF, but also to avoid recombination [58] or increase the Fill Factor [59] using similar strategies that apply in semiconductors technology.

As regard to the IEF driven photoelectric effects and their relation with photostriction, it has been postulated that the converse piezoelectric effect generated by a photo-induced electric field was the main source of deformation cause by light in ferroelectric materials. Recent first-principle calculations have brought further evidence of this scenario [60-62]. However, ref. [62] has shown that some subtleties exist: the piezoelectric effect appears to be the dominant force in directions with large piezoelectric constants and large photo-induced change of polarization (often along or close to the polar axis), meanwhile the photo-induced potential deformation can become sizable in directions perpendicular to the polar axis in lead titanate for instance. Moreover, it was demonstrated that depending on the transition involved, the potential deformation effect can either collaborate, or compete. This means that there is room for improvement and specific design to enhance, or on the contrary impede the photostriction effect, by proper tailoring of the band structure and specific transition selection (by carefully choosing the light polarization for instance).

It is also important to realize that the coupling of light with the lattice dynamics in ferroelectrics offers a myriad of exciting perspective. For instance, it was predicted and demonstrated that coupling of light with IR phonons offered the possibility to reverse polarization, provided strong anharmonic coupling between the IR phonon and soft mode phonons exists [63, 64]. Similarly, ref. [48] shows the existence of large photo-induced strain gradient at the picosecond scale, which could lead to polarization control through the flexoelectric effect. At last, it was shown in ref. [53] that the generation of coherent GHz acoustic phonons by a strong laser pulse allowed for strong and fast acousto-optic sub-bandgap light propagation control, which is a promise for the design of the next generation of ultra-fast acousto-optic devices.

## 3. Concluding Remarks

Perovskite halides based materials [65] are focusing a lot of attention due to its high photovoltaic efficiency, but presenting the important disadvantage of containing toxic materials [66]. Oxide ferroelectrics might be an interesting alternative. The tremendous advances during the last years achieving large efficiencies via the absorption increase (8.1% [13]) and even locally overpassing the Shockley-Queisser-limit [26] are promising in terms of applications. Future prospects for efficiency increase are those that include the optimization of the properties (either "positive or "negative") that make oxide photoferroelectric materials different from the materials currently used in photovoltaics (Figure 1): i) internal fields, ii) high dielectric permittivity, iii) low mobility. There is a lot of room for improvement exploiting the good properties and making less bad the bad ones of photoferroelectrics.

## Acknowledgements

Financial support by the Spanish Government [Projects MAT2014-56063-C2-1-R, MAT2015-73839-JIN, MAT2014-51778-C2-1-R and MAT2014-57960-C3-1-R, and associated FEDER], the Generalitat de Catalunya (2014-SGR-734) is acknowledged. ICMAB-CSIC authors acknowledge financial support from the Spanish Ministry of Economy and Competitiveness, through the "Severo Ochoa" Programme for Centres of Excellence in R&D (SEV- 2015-0496). C. P. and B.D. acknowledge a public grant overseen by the French National Research Agency (ANR) as part of the "Investissements d'Avenir" program (Grant No. ANR-10-LABX-0035, Labex NanoSaclay).

# Corresponding authors

* ignasifinamartinez@gmail.com